\documentclass{article}

\PassOptionsToPackage{numbers,sort&compress}{natbib}
\PassOptionsToPackage{font=footnotesize,labelfont=bf}{caption}

     \usepackage[final]{neurips_2019}

\usepackage[utf8]{inputenc} 
\usepackage[T1]{fontenc}    
\usepackage[hidelinks]{hyperref}       
\usepackage{url, graphicx}            
\usepackage{booktabs}       
\usepackage{amsfonts} 
\usepackage{amsmath}
\usepackage{nicefrac}       
\usepackage{microtype}      
\usepackage{multirow}
\usepackage{afterpage}
\usepackage{subfig}
\usepackage{todonotes}
\usepackage{wrapfig}

\title{Is graph-based feature selection of genes\\ better than random?}

\author{%
Mohammad Hashir\thanks{equal contribution. Correspondence to: Mohammad Hashir<mohammad.hashir.khan@umontreal.ca>, Paul Bertin<bertinpa@mila.quebec>, Joseph Paul Cohen<joseph.paul.cohen@mila.quebec>.}\\
Mila, Université de Montréal\\
\And
Paul Bertin\footnote[1]{} \\
Mila, Université de Montréal\\
\And
Martin Weiss \\
Mila, Université de Montréal\\
\And
Vincent Frappier \\
Mila, Université de Montréal\\
\And
Theodore J. Perkins \\
Ottawa Hospital Research Institute \\
University of Ottawa \\
\And
Geneviève Boucher \\
Institute for Research in Immunology and Cancer \\
Université de Montréal \\
\And
Joseph Paul Cohen \\
Mila, Université de Montréal\\
}

\begin{document}

\maketitle

\begin{abstract}
Gene interaction graphs aim to capture various relationships between genes and represent decades of biology research. When trying to make predictions from genomic data, those graphs could be used to overcome the curse of dimensionality by making machine learning models sparser and more consistent with biological common knowledge. In this work, we focus on assessing whether those graphs capture dependencies seen in gene expression data better than random. We formulate a condition that graphs should satisfy to provide a good prior knowledge and propose to test it using a `Single Gene Inference' (SGI) task. We compare random graphs with seven major gene interaction graphs published by different research groups, aiming to measure the true benefit of using biologically relevant graphs in this context. 
Our analysis finds that dependencies can be captured almost as well at random which suggests that, in terms of gene expression levels, the relevant information about the state of the cell is spread across many genes. Our method is available on github: https://github.com/mila-iqia/gene-graph-conv
\end{abstract}

\section{Introduction}

 Many groups have developed a number of gene-interaction graphs, structuring domain knowledge from different areas of molecular biology \cite{Ogris2018,Warde-Farley2010,Himmelstein2015,Himmelstein2017-vq,Lee2011, Hwang2019-gb,Subramanian2017_s,liu2015regnetwork,Kanehisa2017,Szklarczyk2019}. These graphs can represent any number of different biological, molecular, or phenomenological relationships such as protein-protein interactions, transcriptional regulation, transcriptional co-regulation, co-expression at the mRNA or protein levels, etc. In this work, we focus on gene interaction graphs as a form of prior biological knowledge for machine learning models. 
 
Gene interaction graphs can be used with machine learning algorithms as a proxy for biological intuition to leverage decades of biology research \cite{Zhang2017}. They can act as a biological prior on machine learning techniques to automate feature importance \& selection and help to overcome the curse of dimensionality.
For example, network-based linear regression \cite{Li2008, Min2016} regularizes the weights of a linear model based on the connectivity of the nodes found in an interaction graph. Preliminary work by \citet{Rhee2018} and \citet{Dutil2018} found that the same can be done for non-linear models and remarked that the quality of these graphs may impact their potential in developing general models which would be useful in the majority of tasks where gene expression or single-nucleotide polymorphism data is the input. As these graphs were not developed as an input for machine learning applications, there is value in investigating whether they can aid machine learning algorithms.

We propose a new theoretical interpretation of the approach in \citet{bertin2019analysis} and formulate a condition that a graph should satisfy in order to provide ``good'' prior knowledge for machine learning algorithms. We then test whether this condition holds for random graphs, using a Single Gene Inference approach similar to \cite{bertin2019analysis, Dutil2018, Chen2016, Subramanian2017_s}, and compare random graphs with seven major gene interaction graphs created by different research groups (which we refer to as `curated graphs'), aiming to measure the true benefit of using biologically relevant graphs in this context. Specifically, we construct a single gene inference task and compare the performance of a non-linear model (a multilayer perceptron) using only the first degree neighbours of a gene in the graph against a model that uses the full gene set. 

With this work, we aim at gaining greater insight into the behavior of machine learning pipelines that make use of graph-based prior knowledge in the context of gene expression data. This effort is of primary importance as genomics is a domain where we have relatively limited intuition compared to images or text. Having more interpretable models could provide a ``research gradient'' to biologists allowing them to focus on specific subgroups of genes, which could lead to a fruitful feedback loop between biological experiments and machine learning predictions. Interpreting those models could also help in generating new hypotheses that may be validated with experiments.

\section{What is a ``good'' graph-based prior knowledge?}

For a set of gene expressions $\{g_i\}_{i \in [1...N]}$ where $N$ is the number of genes, the joint probability of the expressions is denoted by $P( g_1, ..., g_N)$. We hypothesize that there is a true causal (directed) graph $G$ that generated this distribution: each gene expression was generated by a function $f$ such that $g_i \doteq f(\mathit{PA}_i, \epsilon)$ where $\mathit{PA}_i$ refers to the set of expressions of the parents of node $i$, and $\epsilon$ is some random noise.

\begin{wrapfigure}{R}{0.4\textwidth}
    \centering
    \includegraphics[width=0.37\columnwidth]{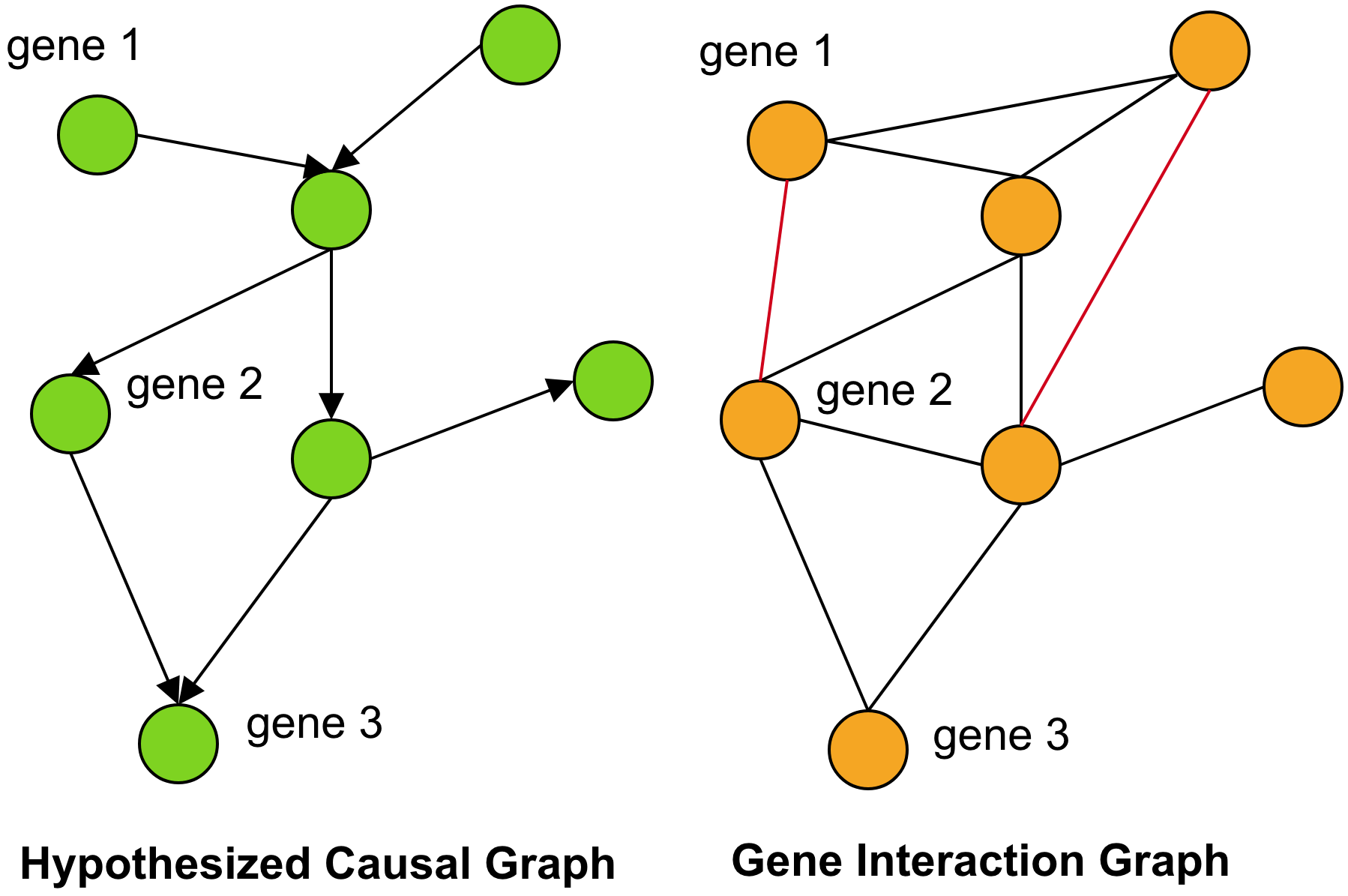}
    \caption{We define a ``good'' prior knowledge as an interaction graph $G$ that covers the moral (undirected) graph equivalent to the hypothesized causal graph, \textit{i.e.} edges in $G$ cover all edges found in the hypothesized \textit{true} causal graph, and parents with a common child are connected in $G$. $G$ could also contain other \textit{spurious} edges (in red) while still satisfying the inclusivity property, as in this example.}
    \label{fig:diagram_graphs}
\end{wrapfigure}

In order to provide ``good'' prior knowledge, we would like our gene interaction graph to be equivalent to the hypothesized \textit{true} causal graph. Figure \ref{fig:diagram_graphs} illustrates the relationship between the two graphs.

\textbf{Definition} \hspace{3pt} A ``good'' prior knowledge is a gene interaction (undirected) graph $G$ which covers the moralized counterpart of the \textit{true} causal graph $H$, \textit{i.e.} if an edge $i \rightarrow j$ exists in $H$, the edge $\{i,j\}$ should exist in $G$, and if two nodes have a common child in $H$, an edge should exist between them in $G$.

\textbf{Inclusivity Property} \hspace{3pt} If an interaction graph $G$ is a ``good'' prior knowledge, then for any gene node $i$, the Markov blanket of $i$ in the \textit{true} causal graph is contained in the set of neighbours of node $i$ in $G$. Equivalently, if $G$ is a ``good'' prior knowledge, the following holds for all gene nodes $i$: 
\begin{equation}
    P(g_i | \overline{g_i}) = P(g_i | neighbours_i) 
    \label{eq:1}
\end{equation}
where $\overline{g_i} = \{g_j\}_{j \ne i}$ is the set that contains every gene expression except the $i^{th}$ one, and $neighbours_i = \{g_j | \exists \text{ edge } \{i, j\} \text{ in graph } G \}$. If the equality (Eq. \ref{eq:1}) holds for a gene $i$, it means that the conditional probability of $g_i$ given all the other genes only depends on the first degree neighbours of $g_i$. 

The inclusivity property does not ensure that the interaction graph has no spurious edges. An edge $\{i, j\}$ in the interaction graph is called spurious if it does not exist in the moralized counterpart of the hypothesized \textit{true} causal graph. Note that the detection of spurious edges is not our main concern as we deal with fairly sparse graphs. The goal of this work is to identify graphs that are sparse while still satisfying the inclusivity property.

\textbf{Method} \hspace{3pt}
As there is no direct way to test the equality (Eq. \ref{eq:1}), we model the conditional probability of the expression of gene $g_i$ given all the other genes with a neural network. This task, predicting a gene expression value given a set of other gene expressions, is similar to the Single Gene Inference task formulated in \citet{Dutil2018}, which was inspired by \cite{Chen2016} and \cite{Subramanian2017_s}. 
For each gene \textit{i}, we train two different models that try to predict its expression level $g_i$. The first model takes all the other genes as input and the second takes only the first degree neighbours as input.
If the equality holds for $g_i$, both models should achieve similar performance and approximate the conditional probability equally well. We can even expect slightly better performance in the second model as signal is supposed to be less noisy and lower dimensional. Conversely, if the equality does not hold for $g_i$, then we expect the second model to achieve poorer performance as it will be provided with incomplete information.

We restrict the prediction of $g_i$ to a binary classification task to simplify interpretation of the results. The alternative is to define a regression task, but depending on the pattern of expression of the gene, its range, its level of noise or any other specificity, the regression metric of a given gene (\textit{e.g.} R-squared) can be arbitrarily high even for a \textit{good} fit. A reduction of performance in some genes can be missed when looking at global statistics of the regression task.
The key point here is that we are interested in aggregated results (\textit{e.g.} mean over all genes) which requires metrics that are comparable between genes. We believe that AUC of a binary classification task matches those requirements. Another reason we use classification is that we are not aiming to precisely model gene expression patterns; rather we intend to compare the two models to know whether the equality (Eq. \ref{eq:1}) holds or not.

Previous work demonstrated that the equality (Eq. \ref{eq:1}) holds for most genes in several graphs \cite{bertin2019analysis}. In this work, we analysed to which extent random graphs allow the inclusivity property to hold, and how they compare to biologically relevant graphs.

\section{Experiments and results}

 \textbf{Datasets} \hspace{3pt} We perform our analysis not only with \textit{healthy} cells (GTEx \cite{lonsdale2013genotype}), but also with \textit{cancerous} cells (TCGA  \cite{CancerGenomeAtlasResearchNetwork2013}) where biological processes might somehow be perturbed, giving us an idea of the usefulness of using gene interaction graphs in different contexts. 
 The TCGA PANCAN database spans multiple tissues and measures 20,530 gene expressions for 10,459 samples; most samples come from cancer biopsies. The GTEx dataset consists of only healthy subjects and has a higher amount of genomic features (34,218 genes) but only 2,921 samples. We normalized both datasets by their respective mean (gene-wise) for our analysis.

 \textbf{Graphs} \hspace{3pt}  We evaluated six graphs covering a variety of relationships in the genome, namely GeneMania \cite{Warde-Farley2010}, RegNetwork \cite{liu2015regnetwork}, Hetionet \cite{Himmelstein2017-vq}, FunCoup \cite{Ogris2018}, HumanNet \cite{Hwang2019-gb} and StringDB \cite{Szklarczyk2019}. For Hetionet, we combined the Interaction, Covariation and Regulation sub-graphs and for StringDB, we evaluated with both the co-expression graph and the entire graph. We also generated a separate graph based on the Landmark genes \cite{Subramanian2017_s} where all the genes in a given dataset are connected to the 978 landmark genes (which are themselves connected together). We did not take into account weighted edges but considered them as present or absent. 
We then generated graphs with a fixed number \textit{n} of randomly sampled neighbours from the set of genes in the dataset. For each target gene in the dataset, we sample \textit{n} other genes from the dataset and connect them to the target gene node to create an \textbf{R-\textit{n} graph}. We created 15 such graphs with \textit{n} varying between 10 and 10,000.

\textbf{Modeling} \hspace{3pt}  In order to predict the over- or under-expression of the target gene compared to its average expression level, we began by binarizing it based on its mean expression. Then, two multi-layer perceptrons (MLPs) were trained to predict the target gene expression using the two types of inputs mentioned above: all the genes (baseline, also referred to as `fully connected' graph) and only the first degree neighbours. The AUC was computed on a test set after training. If the target gene had no neighbours in the graph, an AUC of 0.5 was assigned because the absence of any input features made the prediction a random guess. Further training details are available in the supplementary material. 

We performed these experiments for all the seven curated graphs and 15 R-\textit{n} graphs with both datasets and all the genes in each dataset. We ran three trials for each combination of a gene, graph and dataset and averaged all metrics across the trials for a robust evaluation. For each trial, we used 3000 samples for TCGA and 1500 samples for GTEx with equal splits between the training, testing and validation sets. The data and also the gene neighbours for the R-\textit{n} graphs were randomly sampled for every trial but the data remained the same for every gene regardless of graph. Note that different graphs cover different sets of genes which can bias the aggregated statistics. 

\begin{figure*}[h]
    \centering
        \subfloat[TCGA]{\includegraphics[width=0.5\textwidth]{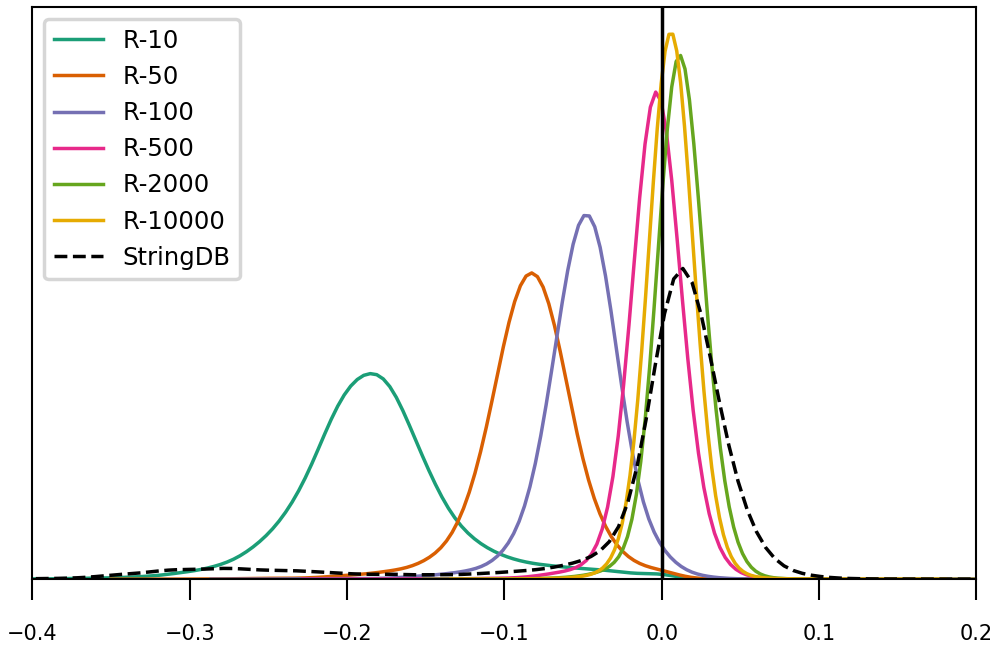}}%
        \subfloat[GTEx]{\includegraphics[width=0.5\textwidth]{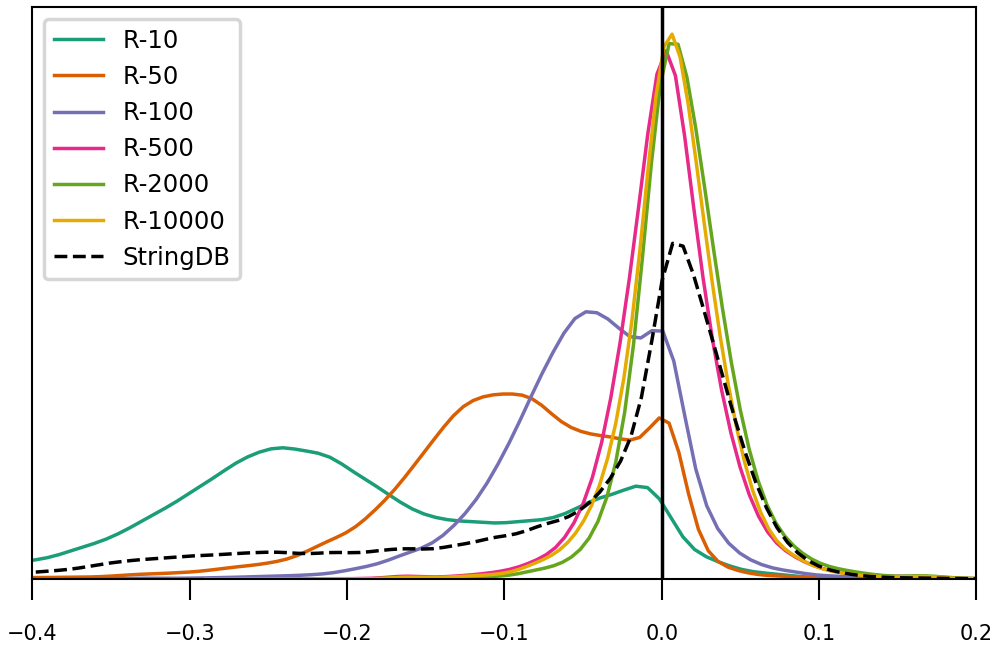}}    \caption{Distribution of AUC improvements over genes in R-\textit{n} graphs. The dashed line represents StringDB, the best performing curated graph. The more towards the right the better the performance.
    }
    \label{fig:rand-diff-auc}
\end{figure*}

\textbf{Results} \hspace{3pt} For a given gene, we define its \textit{AUC improvement} as the difference between the AUC of the model taking first neighbours as input and the AUC of the model taking all other genes as input. For several graphs, we plot the distribution of \textit{AUC improvements} over genes in Figure \ref{fig:rand-diff-auc}. As expected, when the number of random neighbours \textit{n} increased, the feature selection related to the graphs achieved better performance. R-\textit{n} graphs perform on par with or better than the baseline on average for \textit{n} greater than $500$, meaning that using 500 randomly sampled features actually performs on par with or better than using the entire gene set. The standard deviation across trials of the per-gene AUC is $1\mathrm{e}^{-2}$. 

\begin{figure}[h]
    \centering
        \subfloat[TCGA]{\includegraphics[width=0.5\textwidth]{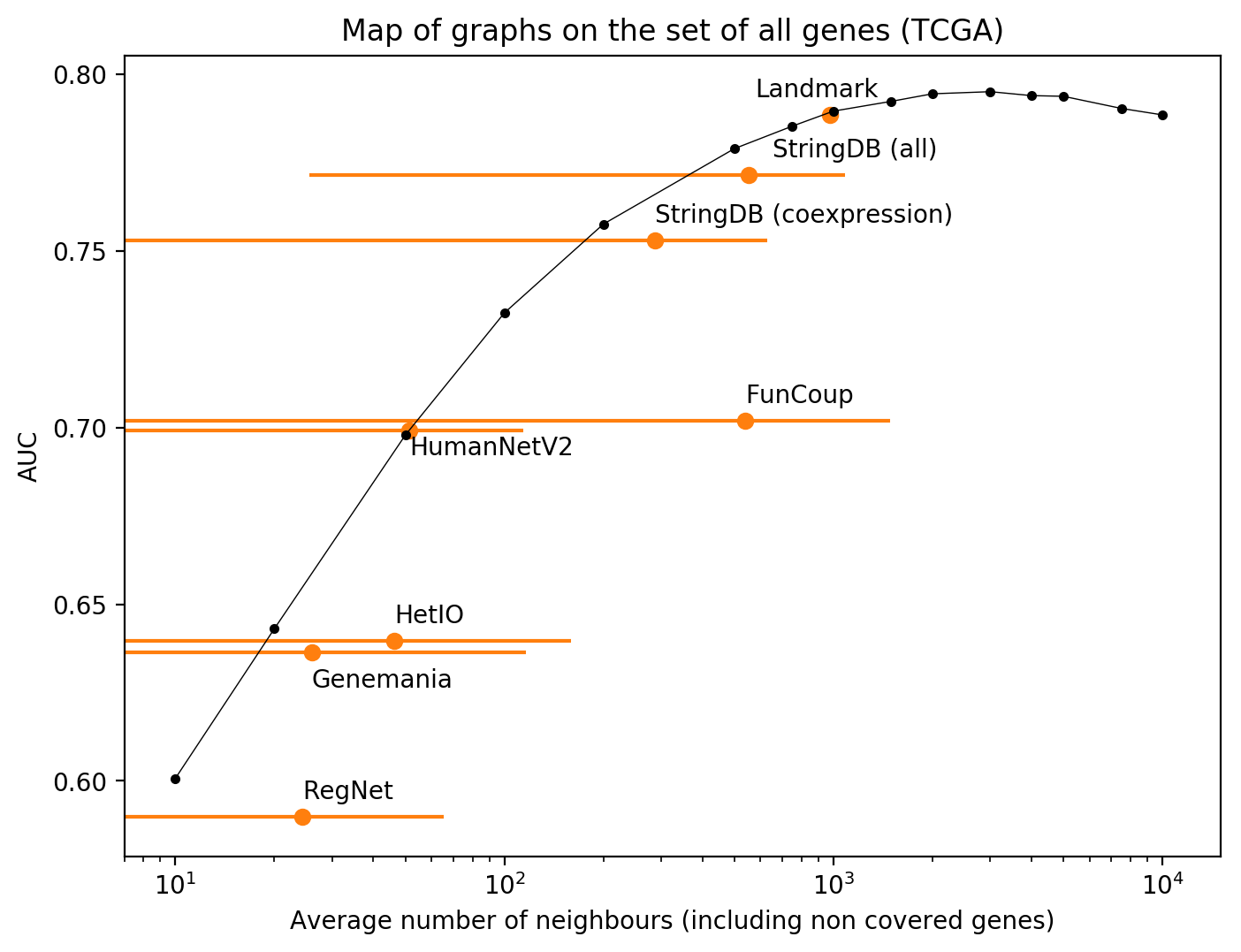}}
        \subfloat[GTEx]{\includegraphics[width=0.5\textwidth]{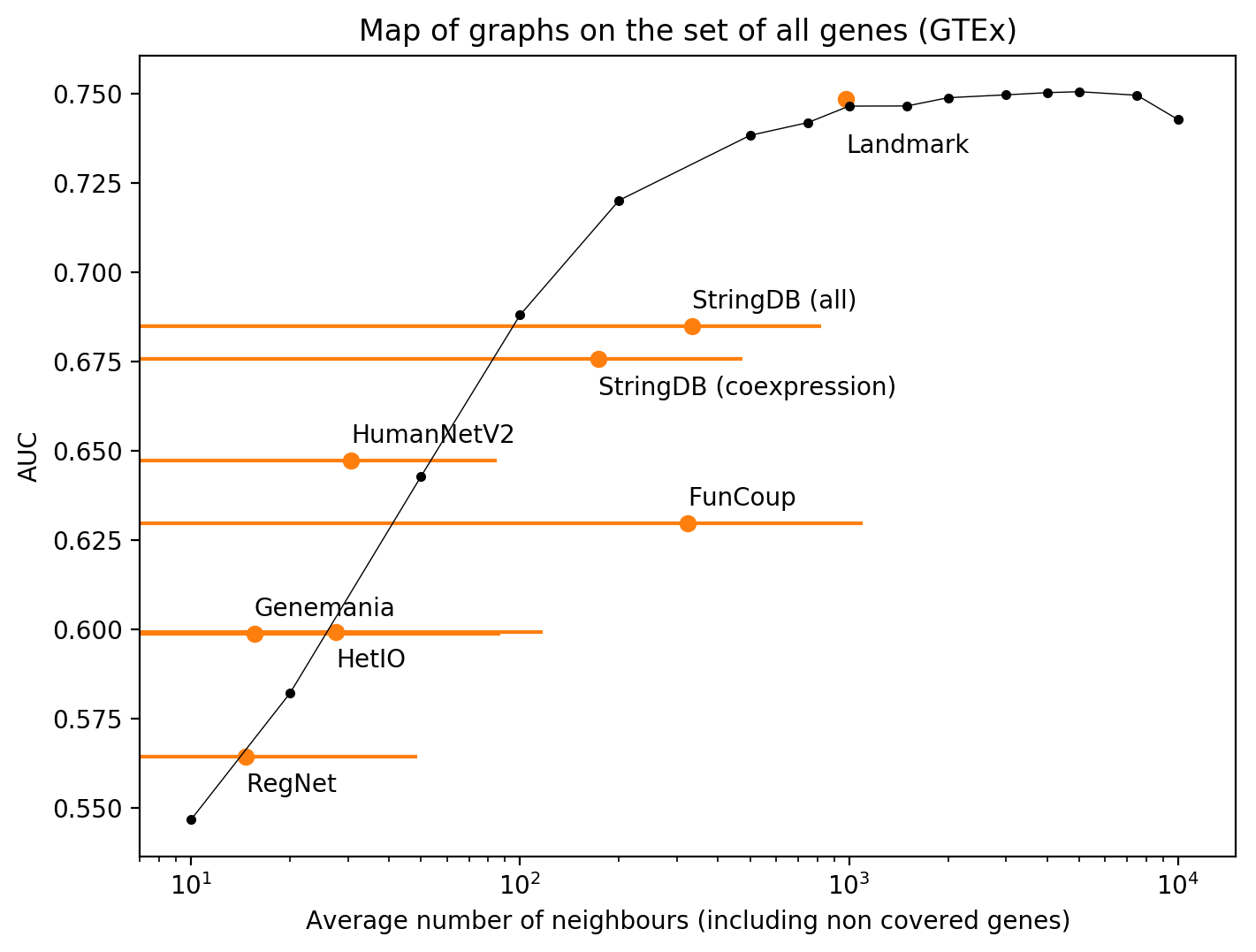}}
        \caption{Mean AUC on all genes as a function of average number of neighbours in the graph (logarithmic scale). The black line represents the average AUCs achieved with the multiple R-\textit{n} graphs and one could think about it as the level of randomness. The orange lines indicate the standard deviation of the number of neighbours for each graph. The standard deviation of AUCs across trials is $2\mathrm{e}^{-4}$ and too small to be plotted.}
    \label{fig:aucVneighbors}
\end{figure}

Figure \ref{fig:aucVneighbors} shows the mean AUC over genes as a function of the average number of neighbours in the graph. The mean AUC is computed over all genes in the dataset. Results on the set of genes which are covered by all graphs are presented in the supplementary material in Figure \ref{fig:aucVneighborsintersection}. The R-\textit{n} graphs are depicted as a black line which could be thought of as the \textit{level of randomness} for the task. Almost all the curated graphs are below the level of randomness except for HumanNetV2 (GTEx), GeneMania (GTEx) and Landmark. Poor performance is in part due to the limited coverage of some of the graphs, in which some genes do not have any neighbours that made the prediction a random guess. On the set of genes covered by all graphs, most curated graphs are above the level of randomness, but only StringDB achieves better performance than the best performing random graph, and even then only by a very small margin ($\leq 5 \mathrm{e}^{-3}$ AUC).
Note that in general, predictive performance was better on GTEx than TCGA as the latter has mostly unhealthy samples where underlying biological processes might be perturbed. 

\section{Conclusion}
We proposed a definition of ``good'' graph-based prior knowledge and derived a property that allows us to test the goodness of the prior knowledge associated with a given graph. Single Gene Inference tasks were used to test whether the property holds for a given graph in the context of predictions on gene expression data.

We then compared existing gene interaction graphs against randomly generated graphs to assess the quality of the prior knowledge they provide. We found that randomly selecting 500 or more genes as neighbours for a target gene can perform on par with or improve over the baseline for most genes in TCGA and GTEx. This means that a random graph R-\textit{500} is a good prior knowledge in which the equality (Eq. \ref{eq:1}) holds for most genes, while being quite sparse. Those results show that the relevant information about the state of the cell is spread across many genes.

Thus, the additive value of using curated graphs to provide prior knowledge appears to be limited. We chose to perform our evaluation on the complete dataset as opposed to the set of genes graphs cover. This is because our goal is to find the most general graph that can be used for a variety of machine learning tasks. Nonetheless, curated graphs could be valuable when one is interested in specific subgroups of genes which have been well studied by biologists. This analysis, as well as the validation of our results in the context of a clinically relevant prediction task are left for future work.


\section*{Acknowledgements}
We thank Francis Dutil and Mandana Samiei for their useful code and comments. This work is partially funded by a grant from the Fonds de Recherche en Sante du Quebec and the Institut de valorisation des donnees (IVADO).  This work utilized the computing facilities managed by Mila, NSERC, Compute Canada, and Calcul Quebec. We also thank NVIDIA for donating a DGX-1 computer used in this work. We thank AcademicTorrents.com for making data available for our research.

\bibliographystyle{plainnat-nopagenum}
\bibliography{main}

\newpage

\section*{Supplementary material}
\subsection{Training details}
We utilized an MLP with a single hidden layer of 16 neurons and ReLU activation functions. The binary cross-entropy loss was used with an Adam optimizer and a learning rate of 0.001 for the FunCoup, Hetionet, and fully-connected graphs and $7\mathrm{e}^{-4}$ for the rest, on all datasets. The weight decay parameter was set to $1\mathrm{e}^{-8}$. These hyperparameters were obtained with a search over the different graphs and datasets, over 20 genes. The MLP achieved slightly better performance than logistic regression ($+5\mathrm{e}^{-3}$ AUC on average over 20 genes). $L_1$ regularization was not used as it did not improve performance.

\subsection{Plots on intersection of genes}
\begin{figure}[!h]
    \centering
        \subfloat[TCGA]{\includegraphics[width=0.5\columnwidth]{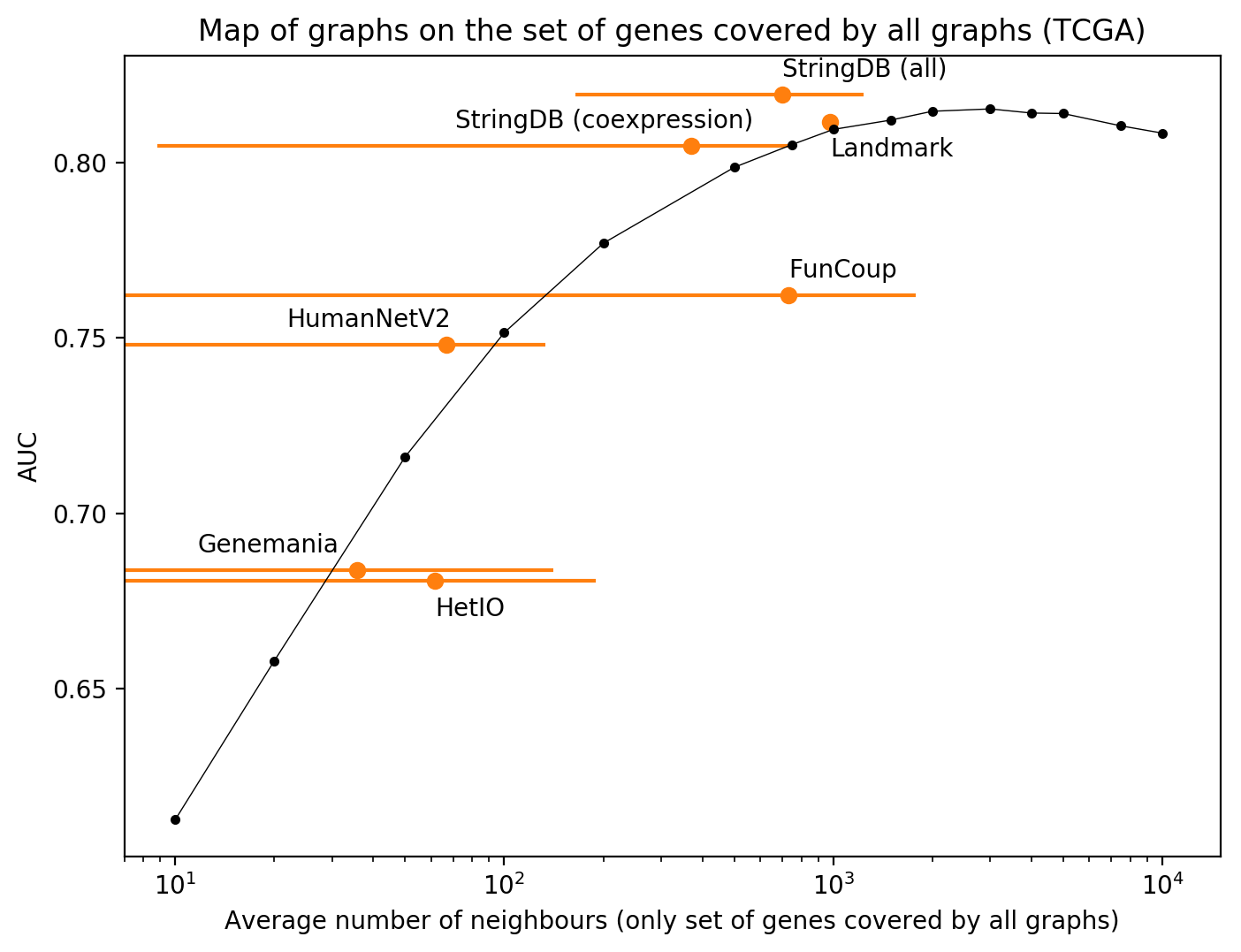}}
        \subfloat[GTEx]{\includegraphics[width=0.5\columnwidth]{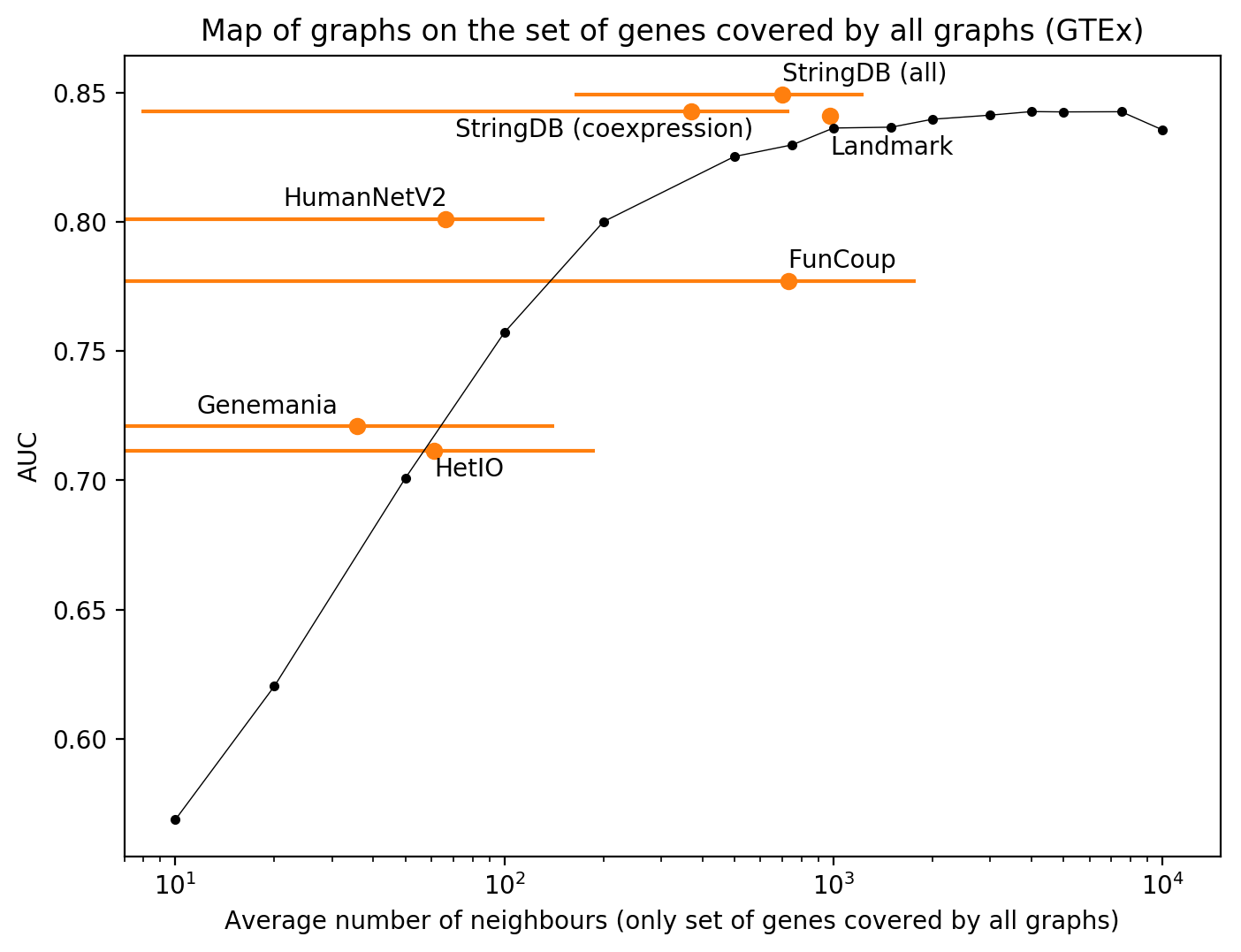}}
        \caption{Mean AUC on the set of genes covered by all graphs (\textit{intersection set}) as a function of average number of neighbours (on the \textit{intersection set}) on a logarithmic scale. RegNet is omitted due to its very small number of gene nodes. The black line represents the average AUCs achieved with the multiple R-\textit{n} graphs and the orange lines indicate the standard deviation of the number of neighbours for each graph. The standard deviation of AUCs across trials is $2\mathrm{e}^{-4}$ and too small to be ploted.}
    \label{fig:aucVneighborsintersection}
\end{figure}
\end{document}